\documentclass[twocolumn,secnumarabic,nofootinbib,tightenlines,nobibnotes,aps,prl,unsortedaddress,superscriptaddress]{revtex4}
\usepackage[usenames,dvipsnames]{color}
\usepackage{pict2e}
\usepackage{mathtools}
\usepackage{hyperref}
\usepackage{graphicx}
\usepackage{subfigure}
\usepackage{bbding}
\usepackage{epstopdf}
\usepackage{graphicx}
\usepackage{amsmath}
\usepackage{amsfonts}
\usepackage{amssymb}
\usepackage{bezier} 
\usepackage{bm}

\begin{document}

\newcommand{\CI}{{\cal I}}
\newcommand{\CT}{{\cal T}}
\newcommand{\mib}{\bm}
\newcommand{\Bk}{{\mib k}}
\newcommand{ \Br}{{\mib r}}
\renewcommand{\Re}{\textsf{Re}}
\renewcommand{\Im}{\textsf{Im}}
\newcommand{\Tr}{\textsf{Tr}}
\newcommand{\id}{\mathbb{I}}
\newcommand{\gz}{G^{(0)}}
\newcommand{\hz}{H^{(0)}}
\newcommand{\grz}{\mathcal{ G}^{(0)}}
\newcommand{\gr}{\mathcal{ G}}
\newcommand{\gc}{\Gamma^{\rm C}}
\newcommand{\ga}{\Gamma^{\rm A}}
\newcommand{\nd}{^{\vphantom{\dagger}}}
\newcommand{\pp}{\text{P}}
\newcommand{\header}[1]{\paragraph{#1}} %for PRL

\title{Stability of Weyl metals under impurity scattering}
\author{Zhoushen Huang}
\affiliation{Department of Physics, University of California at San Diego, CA 92093, USA}
\author{Tanmoy Das}
 
\affiliation{Theoretical Division, Los Alamos National Laboratory, Los Alamos,
NM, 87545, USA}

\author{Alexander V. Balatsky}
 
\affiliation{Theoretical Division, Los Alamos National Laboratory, Los Alamos,
NM, 87545, USA}

\affiliation{Center for Integrated Nanotechnologies, Los Alamos National Laboratory,
Los Alamos, NM 87545, USA}

\affiliation{Nordic Institute for Theoretical Physics (NORDITA), Roslagstullsbacken 23, S-106 91 Stockholm, Sweden}

\author{Daniel P. Arovas}
\affiliation{Department of Physics, University of California at San Diego, CA 92093, USA}
\date{\today}
\begin{abstract}
We investigate the effects of bulk impurities on the electronic spectrum of Weyl semimetals,
a recently identified class of Dirac-type materials.  Using a $T$-matrix approach, we study
resonant scattering due to a localized impurity in tight binding versions of the continuum
models recently discussed by Burkov, Hook, and Balents, describing perturbed four-component
Dirac fermions in the vicinity of a critical point.  The impurity potential is described by
a strength $g$ as well as a matrix structure $\Lambda$. Unlike the case in $d$-wave
superconductors, where a zero energy resonance can always be induced by varying the impurity
scalar and/or magnetic impurity strength, we find that for certain types of impurity ($\Lambda$),
the Weyl node is protected, and that a scalar impurity will induce an intragap resonance over a wide
range of scattering strength. A general framework is developed to address this question, as well as to determine the
dependence of resonance energy on the impurity strength.
\end{abstract}
%\pacs{73.43.Cd}
\pacs{a.b.c}

\maketitle

Pathbreaking discoveries in the areas of graphene and topological insulators have focused
attention on a new class of materials, known as {\it Dirac materials\/} \cite{wehling07,WehlingBalatskyUnpub,HK10,QZ11}.  The hallmark of
these systems is the existence of one or more symmetry-protected Dirac nodes in the electronic band structure, where
the density of states (DOS) becomes vanishingly small, and which have topological implications for the existence of gapless
chiral edge states \cite{Volovik03}.  Such gapless bulk (three-dimensional) band structures have been considered in a
number of recent investigations \cite{Murakami07b,wan11,BB11,XuPRL11,yang11,Heikkila12,Krempa12}.  A more general taxonomy was
advanced by Burkov, Hook, and Balents (BHB) \cite{BHB11}, who described the effects of various homogeneous perturbations on a
$3+1$-dimensional system with a massive Dirac point described by the four-component Hamiltonian
$H_0=\sum_{a=1}^3 k_a\,\Gamma^a + m\,\Gamma^4$
written in terms of Dirac matrices.  This provides a minimal model of a system with both time-reversal
($\CT$) and inversion ($\CI$) symmetries, and with a single tuning parameter $m$ which distinguishes the massless
Dirac point from the normal and topological insulating phases when $m\ne 0$.  BHB found that for sufficiently strong 
${\cal T}$ or ${\cal I}$ breaking perturbations, an intermediate {\it Weyl semimetal\/} phase generally arises, in which
the electronic structure is gapless and characterized by point or line nodes. 

In this paper we consider the effect of localized impurities on the electronic structure in the Weyl semimetal (WS) phase.
In particular, we are interested in the occurrence of resonances in the vicinity of zero energy, where the density of
states vanishes as $\omega^2$ in the WS.  We say that the energy node of a Dirac/Weyl material model $\hz$ is
{\it stable\/} if it does not result in a low energy resonance in the presence of a local impurity
potential $g \Lambda\delta({\mib x})$ with arbitrary $g$, where $g$ is coupling strength and $ \Lambda$ is a matrix
describing the scattering potential.  Such resonances give rise to sharp peaks in the density of states (DOS) which
disrupt the pristine Dirac spectrum \cite{BSB12}.  In addition to its characteristic low energy electronic structure,
the WS is also characterized by its topological properties, which interpolate between those of the non-topological insulator
(NI), where the bulk and edge spectra are both gapped, and the topological insulator (TI) or - when $\CT$ is broken - the
Chern insulator (CI), where the bulk is gapped but gapless edge states participate in quantized surface transport.
Bulk transport consequences for scalar impurities were considered in refs. \cite{BHB11,halasz12}.  

We will focus on the stability of a $4$-band tight binding model of Weyl
materials under local impurity scattering.  One might guess, based on the
more familiar single-band problems, that an impurity resonance or bound
state can be induced at arbitary energy, {\it i.e.\/}, no energy is stable.
This is not true for Weyl materials.  Instead, we find that
stability depends crucially on the type of impurity, which
mathematically can be classified by its commutation relations with the
$\Gamma$ matrices appearing in the local Green's function.  Typically
an impurity is a foreign atom or local crystalline defect in an otherwise pristine material.
Thus, the impurity potential should always involve a local scalar scattering component.
We find that potential scattering (local chemical potential on-site change) will induce an intragap
resonance and therefore will break stability at a single particle level.
We will present a general framework to address the existence of impurity resonances and bound states,
and the dependence of their energies on impurity strength.
We will illustrate this first with a simpler case where both time reversal ($\CT$) and inversion ($\CI$)
symmetries are conserved in the impurity-free system. There, the
(fine-tuned) Dirac node is found to be unstable with $\CI$-even
impurities, but stable with $\CI$-odd ones. The same approach can
be used when $\CT$ and/or $\CI$ are broken by a homogeneous term $\eta\Gamma^{\mu\nu}$,
and the Dirac node is replaced by a pair of Weyl nodes (or a line node) for sufficiently strong $\eta$.
In these cases, stability depends
not only on the impurity type, but also on $\eta$, the strength of the
symmetry breaking term.  Results will be presented for the physically motivated Burkov-Balents
model \cite{BB11} of alternating topological/normal insulator layers in
an external magnetic field ($\eta$) along the stacking direction.  The impurity
classification can be found in Table ~\ref{g12-res}, and the stability
phase diagram in Fig.~\ref{g12-phase}.  The critical field strength
$\eta_{\rm c}$ is found to be related a form of band inversion.

{\it Weyl material model --\/} The models we study are lattice versions of the continuum
models discussed by BHB \cite{BHB11}, and are defined by the following
$\Bk$-space Hamiltonian in the $\Gamma$-matrix basis,
\begin{gather}
  \label{hk0}
  \hz(\Bk) = \xi(\Bk) \,\id + \sum_{i = 1}^3d_i(\Bk) \,  \Gamma^i + m(\Bk) \,
   \Gamma^4 + \eta \,  \Gamma^{\mu\nu}
\end{gather}
where $\xi(\Bk) = -2t\sum_i \cos k_i - \varepsilon_0$, $d_i(\Bk) =
-2t_1 \sin k_i$, $m(\Bk) = -4 t^{\prime} \sum_i(1 - \cos k_i) -
\lambda$, and $\eta$ is taken to be $\Bk$-independent for simplicity.
We adopt the following $\Gamma$ matrix convention: $\Gamma^i =
\tau^x\otimes \sigma^i$ ($i = 1,2,3$), $ \Gamma^4 = \tau^z\otimes
\id$, $ \Gamma^5 = -\tau^y \otimes \id$, and $ \Gamma^{\mu\nu} = i [
\Gamma^{\mu}, \Gamma^{\nu}]/2$. $\tau^i$ and $\sigma^i$ are two sets
of Pauli matrices acting on the orbital and spin degrees of freedom,
respectively. In the model, $t$ and $t^{\prime}$ are hoppings between
same orbitals, and $t_1$ is the (spin-mixing) hopping between
different orbitals. For $\lambda = \eta= 0$, there is a confluence of
the four bands at the single Dirac node $\Bk=0$ with energy $E = -6t -
\varepsilon_0$. Such a Dirac node results from parameter fine-tuning,
as nonzero $\lambda$ will open up a gap. With $\eta =0$, the model is
both time-reversal and inversion symmetric. Time reversal is defined
as $\mathcal{ T} = \mathcal{ K} \mathcal{R}$, where $\mathcal{ K}$ is
complex conjugation and $\mathcal{ R} = \id \otimes i\sigma^y$.
Inversion is defined as $\mathcal{I} = \Gamma^4$ (the two ``orbitals''
being opposite inversion eigenstates). The model will exhibit a
semimetal phase over a range of $\eta$ values.

{\it T-matrix and single impurity -- \/} The effect of localized
impurities can be studied in the standard $T$-matrix formalism \cite{economou06,balatsky06}.
We briefly recall the procedure here to establish notation. The Green's function of the
Hamiltonian $ H = \hz + V$ is $ G = \gz + \gz T \gz$ where $\gz(z) = (z - \hz)^{-1}$ and
$T = V (\id - \gz V)^{-1}$ is the $T$-matrix. Assume $\hz$ is translationally
invariant, and the impurity potential is localized at the spatial
point $r = 0$: $V_{ \Br \Br'} = g\Lambda\delta_{ \Br,0}\delta_{ \Br',0}$.
Then the Green's function connecting $\Br$
and $\Br'$ is $ G_{\Br\Br'} = \gz_{\Br\Br'} + \gz_{\Br0} T\nd_{00}\gz_{0\Br'}$,
where $ T_{00}(z) = ( g^{-1} \Lambda^{-1} - \gz_{00}(z))^{-1}$ is the local
$T$ matrix, and $\gz_{00}(z) = \frac{1}{N}\sum_{\Bk} (z - \hz_{\Bk})^{-1}$ is
the unperturbed local Green's function. Here $\hz_\Bk$ is the Fourier
transform of $\hz$ and $N$ is the number of $\Bk$ points.

\begin{figure}
    \includegraphics[width=0.257\textwidth]{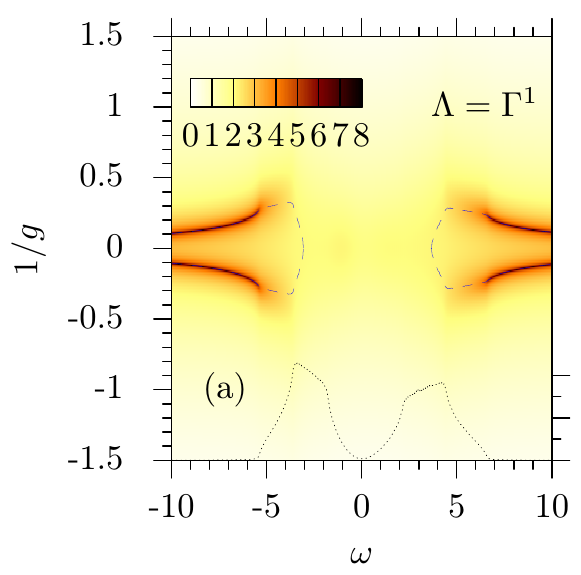}
    \includegraphics[width=0.219\textwidth]{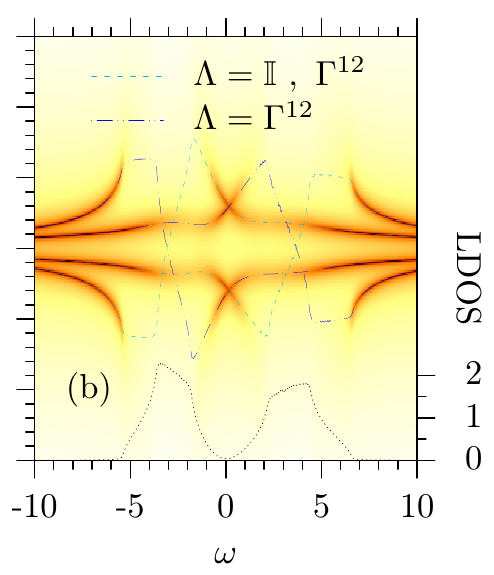}
    \caption{(Color online) Background image: $\log || T_{00}(\omega +
      i\epsilon)||$ for $\mathcal{T}$ and $\mathcal{I}$ symmetric case
      ($\eta = 0$). Darker color corresponds to stronger impurity
      effect. Unperturbed DOS is shown at the bottom.
      Colored lines interpolating the dark curves are obtained by
      replacing $G^{(0)}_{00}$ with $\grz_{00}$ (see text).
      For $\Lambda=\Gamma^1$ (a), the Dirac node is stable.
      Panel (b) shows results for $\Lambda = \Gamma^{12}$ (all curves)
      and for $\Lambda=\id$ (blue dashed curves only); the Dirac node
      is unstable.  Parameters
      used are $t = 0.05, t_1 = -0.5, t^{\prime} = -0.25, \lambda = 0,
      \varepsilon_0 = -0.3$, and lattice size $N_x = N_y = N_z = 50$.
      Spectral broadening $\epsilon$ is set to $0.05$.}
  \label{tnorm-h0}
\end{figure}

\begin{figure}
  \includegraphics[width=0.4\textwidth]{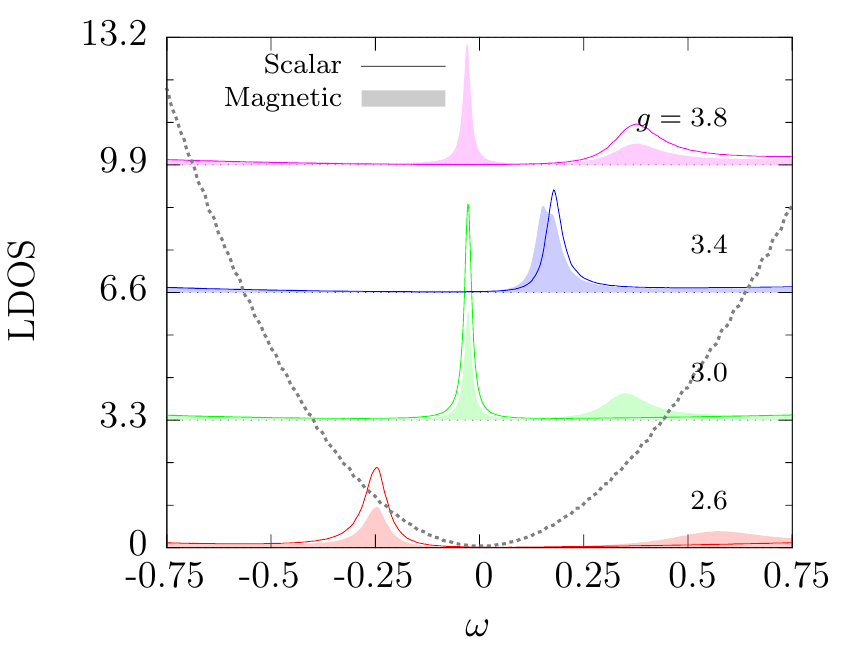}
  \caption{ \label{ldos-single-site} (Color online) Resonance LDOS at
    nearest neighbor site $[x,y,z] = [0,0,1]$ by scalar and magnetic
    impurities. Solid curves: scalar impurity ($ \Lambda = \id$). Shaded
    curves: magnetic scattering ($ \Lambda =  \Gamma^{12}$). LDOS with
    different $g$ couplings are plotted on different base lines. Gray
    dotted curve: unperturbed LDOS rescaled. Same parameters as used
    in Fig.~\ref{tnorm-h0}.}
\end{figure}

The impurity effect at energy $\omega$ can numerically be
characterized by the matrix norm $|| T_{00} || = \sqrt{\sum_a|\lambda_a|^2}$ where $\lambda_a$ are eigenvalues
of $T_{00}(\omega + i0^{+})$. For instance, if $|| T_{00}(\omega + i0^{+})||$ is large,
the local density of states (LDOS) $\rho(\omega) = \rho_{\Br}(\omega) = -\Im\,\Tr\,G_{\Br\Br}(\omega + i0^{+})/\pi$ will in general
deviate significantly from the unperturbed one close to the impurity, yielding a sharp
resonance peak. In Fig.~\ref{tnorm-h0}, we plot $\log ||T_{00}||$ for three impurity forms, showing two qualitatively
different cases: in (a) the Dirac node is stable because $T_{00}$ has no poles near $\omega=0$. For
scalar or purely magnetic impurities (b), the stability of the Dirac node is disrupted by low
energy resonances.  Fig.~\ref{ldos-single-site} plots for
case (b) the LDOS on the nearest neighbor site of the impurity. The positions of the resonance
peaks agree with the large $\log ||T_{00}||$ line in Fig.~\ref{tnorm-h0}(b).

%\begin{widetext}
  \begin{table*}
    % \begin{gather}
    %\label{g12-res}
    \begin{tabular}{| c || c | c | c |}
      \hline
      $ \Lambda$ & $\substack{\text{commutation with}\\  \Gamma^4,\  \Gamma^{12},\ \Gamma^4\Gamma^{12}}$ & $g^{-1}$ yielding $\det( \Lambda/g - \grz_{00}) = 0$ & physical types \\
      \hline\hline
      $\id$, $ \Gamma^4$, $ \Gamma^{12}$, $ \Gamma^{35}$ & $(+,+,+)$ & $s_{\Lambda} /g  = a + s_4 b_1 + s_{12} b_2 + sb_3$ & scalar, $J_z$ and/or layer-swapping\\
      $ \Gamma^{13}$, $ \Gamma^{15}$, $ \Gamma^{23}$, $\Gamma^{25}$ & $(+,-,-)$ & $1/g^2 = (a+s_4b_1)^2 - (b_2 + s_4b_3)^2$ & $J_{\parallel}$ with or without layer-swapping\\
      $ \Gamma^3$, $\Gamma^5$, $\Gamma^{34}$, $\Gamma^{45}$ & $(-,+,-)$ & $1/g^2 = (a+s_{12}b_2)^2 - (b_1 + s_{12}b_3)^2$ & layer-mixing with or without $J_z$\\
      $ \Gamma^1$, $ \Gamma^2$, $ \Gamma^{14}$, $ \Gamma^{24}$ & $(-,-,+)$ & $1/g^2 = (a+sb_3)^2 - (b_1 + sb_2)^2$ & $J_{\parallel}$ with layer-mixing \\
      \hline
    \end{tabular}
    %\notag
    %s_4, s_{12}, s_v = \pm 1\text{: Eigenvalues of } \Gamma^4,  \Gamma^{12} \text{ and }  V\quad , \quad s = \pm 1\ .
    % \end{gather}
    \caption{\label{g12-res} Impurity classification for point-node Weyl semi-metal, with $ \Gamma^{\mu\nu} = \Gamma^{12}$ in eqn.~\ref{hk0}. In the second column, $+$ denotes commute and $-$ anticommute. $a$ and $b_i$ are defined after eqn.~\ref{gz-g12}, principal values are implicitly taken. $s_4, s_{12}$ and $s_\Lambda$ are eigenvalues of $ \Gamma^4, \Gamma^{12}$ and $ \Lambda$, respectively, and take the value $\pm 1$. $s \equiv s_4 s_{12}$. In the last column, $J_z$ refers to magnetic impurities in the $z$ (stacking) direction, and $J_{\parallel}$ those in the $xy$ (layer) direction. See Fig.~\ref{g12-phase} for stability phase diagram of the last three classes.}
  \end{table*}
%\end{widetext}

{\it Bound states and resonances -- \/} Bound states appear as poles in $G(\omega)$, which is
to say zeros of $T_{00}^{-1}(\omega)=g^{-1} \Lambda^{-1} - \gz_{00}(\omega)$.  This may occur when
$\omega$ lies outside the bulk bands.  Resonances are solutions where $\omega$ is extended
to have a finite imaginary part, which is inversely proportional to the resonance lifetime.
In solving numerically for the resonances, it is convenient to define $\grz_{00}(\omega)=
{1\over 2}\big(\gz_{00}(\omega+i\epsilon) + [\gz_{00}(\omega+i\epsilon)]^\dagger\big)$, where $\epsilon\to 0$ is
always greater than the energy spacing between consecutive quantized levels.
We then seek solutions to $u_a(\omega)=g^{-1}$, where $\{u_a(\omega)\}$ are the eigenvalues of $\grz_{00} \Lambda$.
This prescription works well when the bulk DOS is small, as we show in Fig. 1, and is exact when the DOS vanishes.
Thus, if, for a given real value of $\omega$, at least one of the eigenvalues $\{u_a(\omega)\}$ is real, {\it and\/}
if the bulk DOS is small, then a bound state or resonance exists for coupling $g=1/u_a(\omega)$.
If all $u_a(\omega)$ have imaginary part, then $\omega$ is a stable energy.  An immediate consequence
is that an impurity with $\big[\Lambda,\grz_{00}\big]=0$ will for some $g$ disrupt the stability at $\omega$, because product of
commuting Hermitian matrices has real eigenvalues. Single-band problems fall in this category as $\Lambda = \id$.

{\it Dirac node -- \/}  To illustrate the stability criteria, we consider first the case with $\eta = 0$, where both
$\CI$ and $\CT$ symmetries are present.  Inverting Eq. \ref{hk0}, one finds that the only $\Gamma$ matrix appearing in
$\gz_{00}$ is $ \Gamma^4$, which is also the inversion operator,
\begin{gather}
  \label{gz00-weyl}
  \gz_{00}(\omega) = \frac{1}{N} \sum_\Bk \frac{1}{\omega - \hz(\Bk)} =
  a(\omega) \, \id + b(\omega)  \, \Gamma^4
\end{gather}
with $a = {1\over N}\sum_k[\omega - \xi(\Bk)]/D(\Bk)$, $b = {1\over N}\sum_k m(\Bk)/D(\Bk)$, and
$D(\Bk) = [\omega - \xi(\Bk)]^2 - \sum_i d_i^2(\Bk) - m^2(\Bk)$.  In the
following, we will restrict to impurities where $\Lambda$ is either the identity or a single $\Gamma$-matrix;
linear combinations thereof can be analyzed in the same fashion.  $\Lambda$ belongs to one of two
classes: $\gc$ which commutes with $\Gamma^4$, and $\ga$ which
anticommutes. Following our general criteria, we solve for real
eigenvalues of $\grz_{00}  \Lambda$. The commuting case has already been
discussed, thus inversion-even impurities may disrupt stability at
any energy, including the Dirac node.  As an example, the purely
magnetic impurity $ \Lambda =  \Gamma^{12} = \id \otimes \sigma^z$ is
shown in Fig.~\ref{tnorm-h0}(b), where dashed lines are the
eigenvalues $u_a$.  The corresponding LDOS on the nearest neighbor site
to the impurity is plotted in Fig.~\ref{ldos-single-site} for various
$g$ which induces resonance around the Dirac node. For inversion-odd
impurities, we have $\grz_{00}\ga = a\,\ga + b \,\Gamma^4 \ga$, and
$u(\omega) = \pm\sqrt{a(\omega)^2 - b(\omega)^2}$,
where principal values of $a$ and $b$ are implicitly taken, as will be
all such coefficients in the rest of the paper. Reality of
$g^{-1} = u$ then requires $|a(\omega)| > |b(\omega)|$. A
representative case with $ \Lambda =  \Gamma^1 = \tau^x \otimes
\sigma^x$ is shown in Fig.~\ref{tnorm-h0}(b), where real $u$ are the
dashed blue lines.

The region between the two LDOS ``towers'' is stable in
Fig.~\ref{tnorm-h0}(b). This in fact is true for all inversion-odd
impurities and is not accidental.  This follows analytically from the
\emph{band center approximation} (BCA) in which the local Green's
function, $\grz_{00}$, is replaced with the Green's function of the local
Hamiltonian, $\bar G(\omega) \equiv (\omega - \hz_{00})^{-1}$, where
$\hz_{00} = \frac{1}{N} \sum_\Bk \hz(\Bk)$. It is easy to verify that
$\hz_{00}$ has eigenvalues $\Omega_\pm = -\varepsilon_0 \pm\delta$, with $\delta = |12t^{\prime} + \lambda|$,
which are doubly degenerate due to inversion symmetry.  We shall call
$\Omega_\pm$ band centers as they represent in some sense the average
position of the bands.  The eigenvalues of $\bar G(\omega)$ are thus
$1/(\omega - \Omega_\pm)$. Comparing with eqn.~\ref{gz00-weyl}, we find the
BCA for $a$ and $b$ as $a^{BCA} = \frac{\omega +\varepsilon}{(\omega +\varepsilon )^2 - \delta^2}$
and $b^{BCA} = -\frac{\delta}{(\omega +\varepsilon)^2 -   \delta^2}$.  Invoking our earlier results,
$g = 1/u$ is imaginary for $\omega \in [\Omega_-, \Omega_+]$.  Thus the Dirac node,
being in between the band centers in general, is stable with $\CI$-odd impurities.

{\it Weyl semimetal phase -- \/} BHB \cite{BHB11} found a WS phase
intervenes when $\CT$ or $\CI$ is broken sufficiently strongly.
Depending on $ \Gamma^{\mu\nu}$ in Eq. \ref{hk0}, degeneracy of the
two central bands may occur either at discrete points or along a line
in the Brillouin zone. We now consider the effect of local impurity on
the WS phase induced by the $\CT$-breaking term $\Gamma^{\mu\nu} =
\Gamma^{12} = \id \otimes \sigma^z$. A complete classification will be
presented elsewhere \cite{HABunp}. The Hamiltonian Eq. \ref{hk0} can
be used to model an alternating stack of topological insulator and
normal insulator layers immersed in a magnetic field along the
stacking direction (the $ \Gamma^{12}$ term), as proposed in
Ref.~\onlinecite{BB11}. The local Green's function is
\begin{gather}
  \label{gz-g12}
  \gz_{00}(\omega) = a(\omega) \, \id + b_1(\omega) \, \Gamma^4 + b_2(\omega) \, \Gamma^{12} +
  b_3(\omega) \, \Gamma^4 \,\Gamma^{12}
\end{gather}
where all coefficients can be obtained analytically \cite{HABunp}.
Note that all four terms in Eq. \ref{gz-g12} mutually commute.

\begin{figure}
  \centering
  \includegraphics[width=0.47\textwidth]{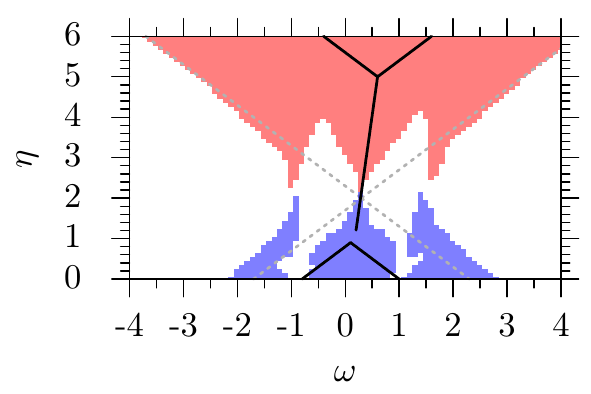}
  \caption{(Color online) Stability of $\mathcal{T}$-breaking Weyl
    semimetal. The vertical axis $\eta$ is the strength of the
    $\mathcal{T}$-breaking term. Red: stable zone for impurity classes
    $(+,-,-)$ and $(-,-,+)$. Blue: stable zone for impurity classes
    $(-,+,-)$ and $(-,-,+)$. See Table \ref{g12-res} for impurity
    classification. Solid black lines mark the two band edges bounding
    the central gap. They touch from around $\eta = 1$ to $5$,
    corresponding to the Weyl semimetal phase. The black lines are
    broken around $\eta = 1$ due to the closing of the indirect gap.
    Dotted gray lines are the stable zone boundaries given by the
    band center approximation. Parameters used are $t = 0.05, t_1 =
    -0.5, t^{\prime} = -0.25, \lambda = 1, \varepsilon_0 = -0.3$, on a
    lattice of $N_x = N_y = N_z = 50$. }
  \label{g12-phase}
\end{figure}

Following the general stability criteria, we find that impurities fall
into four classes, based on the commutation relations of $\Lambda$
with the three $\Gamma$ matrices in eqn.~\ref{gz-g12}. The results are
summarized in Table \ref{g12-res}. The fully commuting class $(+,+,+)$
always yields a real $g$ solution for any $\omega$ and $\eta$. For the
other three classes, there are stable zones in the $\omega$-$\eta$
plane where $g$ is imaginary. A phase diagram is shown in
Fig.~\ref{g12-phase}, where red is the stable zone of the class
$(+,-,-)$, blue of $(-,+,-)$, and $(-,-,+)$ is stable in both red and
blue zones. The general shape of the zone boundary can be understood
by invoking the band center approximation again, \emph{viz.}
$\grz_{00}(\omega) \simeq \big(\omega - \hz_{00}\big)^{-1}$. This
yields the two dotted lines in Fig.~\ref{gz-g12} given by
\begin{gather}
  \omega = \Omega_\pm = -\varepsilon_0 \mp \big(|\delta| - |\eta|\big) \ ,
  \end{gather}
which are nothing but the central two of the band centers (eigenvalues of $\hz_{00}$).
A band center inversion occurs when $\eta_{\rm c}=\pm\delta$,
at which point $\Omega_+ = \Omega_-$. It may be viewed as a
stability critical point (critical external magnetic field) in the
sense that the stable zones of the classes $(+,-,-)$ and $(-,+,-)$ are
each restricted to only one side of $\eta_{\rm c}$. One may think of
impurity levels as forming an energy band parameterized by the
strength $g$, then stable zones in the phase diagram are gaps of the
``impurity band'', and $\eta_{\rm c}$ marks a transition between gapless and
gapped phases. This is reminiscent of the Bloch band inversion
associated with the topological phase transition in Chern insulators.

\begin{figure}
  \centering
  \includegraphics[width=0.5\textwidth]{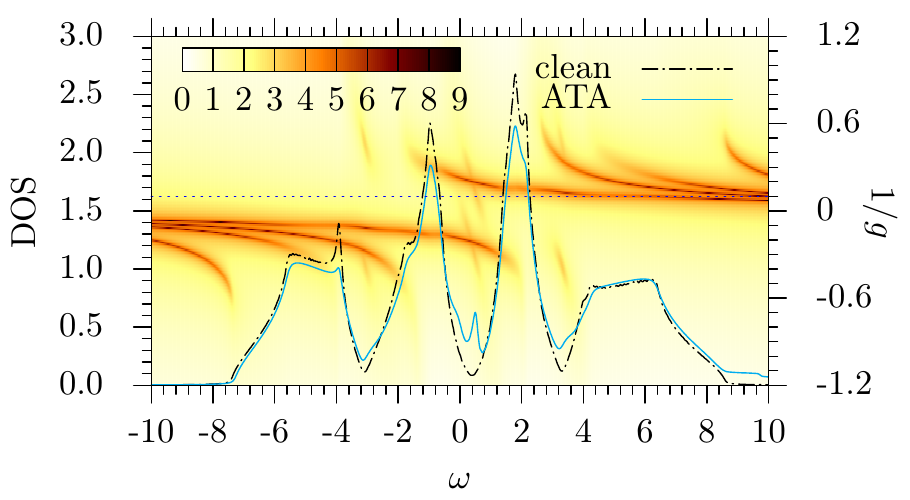}
  \caption{(Color online) Average DOS using ATA for $\Gamma^{\mu\nu} =
    \Gamma^{12}$ (external magnetic field along $\hat z$) and scalar
    impurity $\Lambda = \id$. Impurity concentration is $c = 10\%$ and
    impurity strengths are uniformly distributed in $g \in (0,10]$.
    Solid blue line: ATA result. Dash-dot black line: DOS of clean
    system. Background color image: $\log||T_{00}||$ of single
    impurity as a function of $\omega$ and $g^{-1}$, similar to
    Fig.~\ref{tnorm-h0}. Dotted horizontal line marks maximum $g$ used
    in ATA ($g^{-1} = 0.1$), the ATA DOS is significantly
    enhanced in the range of $\omega$ in which the high
    $\log||T_{00}||$ lines exist \emph{above} $g^{-1} = 0.1$.
    $\eta$ is set to $3$, putting the clean system in the WS phase
    (Fig.~\ref{g12-phase}). Other parameters used are the same as in
    Fig.~\ref{g12-phase}.}
  \label{g12-ata}
\end{figure}

{\it Average $T$-matrix approximation (ATA)} -- The resonance/bound
states as discussed before will broaden into a ``band'' when multiple
impurities are present. Here we consider an ensemble of local
impurities of the same matrix form, with a homogeneous spatial
concentration $c$ and a distribution of strength $f(g)$. We find that
average DOS is enhanced at resonance energy $\omega_{\rm res}(g)$
obtained from a single impurity when $f(g)$ is relatively high. In the
ATA formalism \cite{economou06,balatsky06}, translational invariance
is restored after statistical averaging, and the effect of the
impurity ensemble is captured by the local self energy $\Sigma_{\rm
  loc} = c\,\langle T_{00}\rangle [1 + c\,\gz_{00}\langle
T_{00}\rangle]^{-1}$, where $\langle\cdots \rangle$ denotes averaging
over $f(g)$. The local Green's function is $G_{\rm loc}(z) =
\frac{1}{N}\sum_k 1/(z - \hz(k) - \Sigma_{\rm loc})$ and the average
DOS is $\rho(\omega) = -\Im\,\Tr\, G_{\rm loc}(\omega + i0^{+})/\pi$. In
Fig.~\ref{g12-ata}, we show the average DOS of a sample immersed in an
external magnetic field ($\Gamma^{\mu\nu} = \Gamma^{12}$) with
magnetic impurities breaking local inversion symmetry ($\Lambda =
\Gamma^3$). The impurity strengths are uniformly distributed in $g \in
(0,10]$. The difference in DOS with the clean system agrees with the
single impurity $||T_{00}||$ (shown as colored background).

\header{Acknowledgement} We are grateful to A. Vishwanath, R. Biswas,
and A. Black-Shaffer for useful discussions. This work was supported
in part by the NSF through grant DMR-1007028. Work at LANL was
supported by US DoE Basic Energy Sciences and in part by the Center
for Integrated Nanotechnologies, operated by LANS, LLC, for the
National Nuclear Security Administration of the U.S. Department of
Energy under contract DE-AC52-06NA25396. Work at Nordita was supported
by ERC and VR.

\bibliographystyle{apsrev-no-url} \bibliography{weyl-imp}
%\clearpage
%\appendix
%some appendix..
\end{document}